# Direct laser acceleration of electrons in high-Z gas target and effect of threshold plasma density on electron beam generation

D. Hazra[1,*], A. Moorti[1,2,†], S. Mishra[2], A. Upadhyay[2], and J. A. Chakera[1,2]

[1]Homi Bhabha National Institute, Training School Complex, Anushakti Nagar, Mumbai-400094, India

[2]Laser Plasma Division, Raja Ramanna Centre for Advanced Technology, Indore-452013, India

[*]Email: dhazraphys@gmail.com

[†]Corresponding Author: moorti@rrcat.gov.in

**Abstract:**

An experimental study of laser driven electron acceleration in $N_2$ and $N_2$-He mixed gas-jet target using laser pulses of duration ~60-70 fs is presented. Generation of relativistic electron beam with quasi-thermal spectra was observed at a threshold density of ~$1.6 \times 10^{18}$ cm$^{-3}$ in case of pure $N_2$, and the threshold density was found to increase with increasing doping concentration of He. At an optimum fraction of 50% of He in $N_2$, generation of quasi-monoenergetic electron beams was observed at a comparatively higher threshold density of ~$2 \times 10^{18}$ cm$^{-3}$, with an average peak energy of ~168 MeV, average energy spread of ~21%, and average total beam charge of ~220 pC. Electron acceleration could be attributed to the direct laser acceleration as well as the hybrid mechanism. Observation of an optimum fraction of He in $N_2$ (in turn threshold plasma density) for comparatively better quality electron beam generation could be understood in terms of the plasma density dependent variation in the dephasing rate of electrons with respect to transverse oscillating laser field. Results are also supported by the 2D PIC simulations performed using code EPOCH.

2Keywords: Laser Plasma Acceleration, Laser Wakefield Acceleration, Direct Laser Acceleration, Ionization Induced Injection

# 1. Introduction:

Since the proposition of laser wakefield acceleration (LWFA) mechanism in 1979 [1], it has been studied extensively over years [2-4] and experimental conditions for generation of quasi-monoenergetic electron beams have been identified [5-7]. At the same time, in intense laser plasma interaction different acceleration regimes could be applicable in different laser and plasma conditions [2-4]. Laser wakefield acceleration (LWFA), where electrons gain energy from the wakefield driven by intense laser pulse, was found to be applicable in following conditions: 1) for short laser pulse duration (L<$\lambda_p$, where L is the laser pulse length and $\lambda_p$ is the plasma wavelength), where bubble acceleration mechanism [8-9] was identified, and 2) for long laser pulse duration (L>$\lambda_p$ and L>>$\lambda_p$), self-modulated LWFA (SM-LWFA) mechanism was found to be applicable [10]. Bubble regime of acceleration has led to generation of GeV class high quality quasi-monoenergetic electron beams [11-18]. In the case of SM-LWFA, earlier using several hundreds of fs laser pulses electron beams with broad spectrum [19-22], and later quasi-monoenergetic spectrum were observed using few tens of fs laser pulses [23-28]. In the experimental conditions similar to SM-LWFA i.e. for L>$\lambda_p$, another acceleration mechanism of direct laser acceleration (DLA) of electrons has been identified where electrons gain energy directly from the transverse laser field itself while making transverse oscillations in the plasma channel created by laser [29-33].

For laser plasma acceleration experiments low-Z gases such as $H_2$ and He are preferred as gas-jet targets. Since, $H_2$ is highly reactive and explosive in nature, mostly He is used. However, He being a rare gas is expensive. In low-Z gas targets complete ionization and plasma formation takes place at the foot of the laser pulse and main laser pulse interact with the fully ionized plasma therefore ionization effects on laser propagation could be avoided. However, for injection of electrons in the wakefield one has to rely on the wave breaking process [34]. For high power short laser pulses wave breaking, leading to self-injection and acceleration, generally occurs at a threshold laser intensity of $a_0 \sim 4$ [35], and with longer laser pulses working in SM-LWFA regime even requires higher density to facilitate modulation of the laser pulse [21-24]. Being a highly non-linear process, self-injection and acceleration of electrons through wave breaking mostly leads to generation of unstable electron beams.

Pure high Z gases are not preferred for laser plasma acceleration experiments primarily due to inherent drawback of ionization induced defocusing of the laser pulse [36]. Still, electron acceleration in high Z gases e.g. $N_2$ and Ar etc. have been explored for comparison purpose and particularly $N_2$ which is readily available and cheap [37-48]. In few studies generation of quasi-monoenergetic electron beams with peak energy in the range of few hundreds of MeV has been demonstrated [43-46]. High Z gas targets like $N_2$ has another advantage of supporting ionization induced injection mechanism of electrons [49].

In order to have the advantage of ionization induced injection mixed gas targets were also used where few percent (0.1-10%) of high Z gas (such as $N_2$, $CO_2$, $O_2$ etc.) is mixed in low Z gas (such as $H_2$, He) and generation of few hundred MeV to GeV electron beams have been reported [50-52]. However, due to continuous injection of electrons in the wakefield mostly generation of continuous broad electron spectrum has been observed [50-52]. Several techniques to improve the beam quality have been adopted e.g. the use of dual-stage targets of injector and accelerator [53,54], a combination of ionization and shock-front injection [55], ionization induced injection in a density down ramp [56] where quasi-monoenergetic electron beams have been observed.

Particularly, self-truncated ionization induced injection (STII) scheme [57] is of interest and using that generation of ~1 GeV electron beam of few pC to few tens of pC charge and energy spread of few % was demonstrated [58]. Recently, relying on the STII technique, experimental study to increase the beam charge up to beam loading, while retaining the energy and quasi-monoenergetic feature of the electron beam, in a mixed gas target of He+0.2-1.2%$N_2$ was investigated and electron beam with peak energy of ~250 MeV with energy spread of ~15% and higher charge of ~250-300 pC was demonstrated [59, 60].

As discussed earlier, both wakefield and DLA regimes could be applicable in laser plasma acceleration depending upon the laser and plasma parameters. He gas-jet targets have been extensively studied with short laser pulses operating in the wakefield regime [11-28]. In case of $N_2$ targets, experiments have been performed both in the SM-LWFA [37-43] and LWFA regime [44-48]. In case of mixed gas targets, electron acceleration has been mostly attributed to the wakefield mechanism [50-56, 58-60]. However, in most of the experiments role of DLA in acceleration has not been discussed, as particularly for $L>\lambda_p$ when the laser pulse overlaps with the injected electrons, contribution of DLA cannot be ignored. Electron acceleration by DLA mechanism has been demonstrated in few studies with longer laser pulse duration of few hundreds of fs in both He [30-32] and $N_2$ [37] gas-jet target. Even in the case of shorter laser pulse duration of few tens of fs ($L \geq \lambda_p$) role of DLA along with wakefield have also been considered in pure He target [61, 62]. Pure DLA acceleration of electrons has also been considered in Ar gas-jet targets [63-65]. Recently, Shaw *et al.*, [66] showed hybrid (wakefield + DLA) acceleration mechanism in a He + $N_2$ mixed gas target in the regime $L>\lambda_p$. Further, through detailed PIC simulations it was shown that DLA can double the energy of accelerating electrons over wakefield [67-70]. In our recent investigations also, we identified pure DLA and hybrid regime of acceleration in pure $N_2$ and He + $N_2$ mixed gas target for similar condition of $L>\lambda_p$ [71]. Role of DLA in laser plasma acceleration is a subject of current investigation. DLA regime of acceleration could be favorable for generating higher-energy betatron radiation (x-rays) [72, 73] compared to the wakefield mechanism as in the former case electrons experience a larger transverse oscillation amplitude in the plasma channel. Further, it has been observed through simulations that ionization induced injection mechanism also helps in DLA by giving an additional transverse electron momentum to the electrons [69, 70]. As there are only limited experimental reports on the role of DLA in laser plasma acceleration, further investigations both in high-Z and mixed gas targets would be of interest.

In this report, we present an experimental study of electron acceleration in $N_2$ and $N_2$ - He mixed gas target. Experiments were performed using Ti: Sapphire laser pulses of duration ~60-70 fs (intensity: ~3.7-4.3×10$^{19}$ W/cm$^2$, $a_0$~4.2-4.5) interacting with gas-jet nozzle of 4 mm length. Variation in doping percentage of He in $N_2$ from 0% (i.e. a pure $N_2$ target) to 98% leads to increase in threshold density for electron beam generation from ~1.6×10$^{18}$ cm$^{-3}$ (for pure $N_2$ target) to ~4.3×10$^{18}$ cm$^{-3}$ (for 2% $N_2$ + 98% He target). With pure $N_2$ gas-jet target electron beams (divergence ~6.2 mrad) with average peak energies of ~117 MeV, average maximum energy of ~151 MeV were observed at a threshold density of ~1.6×10$^{18}$ cm$^{-3}$. The total beam charge was in the range of ~65-174 pC. The overall energy spectrum was quasi-thermal. Further investigations with varying percentage of He in $N_2$ showed an optimum composition of $N_2$ + 50% He where generation of quasi-monoenergetic electron beam was observed at a threshold density of ~2×10$^{18}$ cm$^{-3}$. Stable generation of collimated (divergence ~7.5 mrad), quasi-monoenergetic electron beams with higher average peak energy of ~168 MeV, and maximum energy of ~206 MeV and average energy spread of ~21 % were observed. Total beam charge also increased to ~140-300 pC. For higher concentration of He, at increased threshold density, decrease in electron energy and probability of generation of QM beams were observed. Both pure DLA and hybrid (DLA + wakefield) regimes of acceleration of electrons have been identified in the given experimental conditions. Further, effect of varying doping concentration of He (i.e. change in threshold plasma density) on electron beam properties has been explained in terms of varying rate of change of dephasing of electrons with the laser field. Further, 2D PIC simulations using code EPOCH [74] were performed which clearly showed the role of DLA in the present experimental conditions. Also, the effect of plasma density on dephasing of electrons in DLA could be inferred. In summary, by varying the doping concentration of He in $N_2$ mixed gas target, in a single experimental set up, we could control the threshold density for electron acceleration and hence dephasing rate of electrons with the laser field which in turn also control the electron beam properties. Such a study using a gas-jet target of He and high-Z $N_2$ composition with a dominant role of DLA has not been reported so far.

**2. Experimental set up:**

Fig. 1 shows the schematic of the experimental set up. Ti:sapphire laser pulse of duration ~60 - 70 fs was focused to a spot of ~6.5 × 6.5 μm$^2$ (radius at 1/e$^2$) using f/5 optics on a gas-jet target. Considering 38% of total energy inside focal spot, the laser pulse provides a total power (P) of ~17.4 - 20.3 TW and intensity of ~3.7 - 4.3×10$^{19}$ W/cm$^2$ (a$_0$ ~ 4.2 - 4.5). A well characterized slit nozzle of dimension 1.2 mm × 4 mm was used, and laser was propagated along 4 mm length [75]. High-Z gas target of pure $N_2$ and mixed gas target of $N_2$ + varying % of He was used. Plasma density was varied in the range of ~1.6 - 4.3×10$^{18}$ cm$^{-3}$). The plasma density for He and $N_2$ was estimated considering $2^+$ and $5^+$ ionization state respectively for the laser intensity used. Electron beam spectrum was recorded using a magnetic spectrograph consisting of C-shaped rectangular magnet with peak magnetic field strength of ~0.9 T (dimensions 10×10×20 cm: pole gap 15 mm) covering a broad energy range of ~15 MeV to 1 GeV, with a resolution of ~8% at 100 MeV and 25% at 300 MeV (for 10 mrad divergence beam). Three phosphor screens were used for simultaneous recording of electron beam profile and its spectrum. Phosphor-1 (Lanex-regular) was placed before the magnet and at a distance of 39 cm from the gas-jet for recording of electron beam profile. Phosphor-2 (DRZ-High) was placed at a distance of 70 cm from the source in the laser beam direction after Phosphor-1 for recording electron beam profile as well as spectrum. Magnet was inserted in between Phosphor-1 and Phosphor-2 for recording spectrum on Phosphor-2, which covered energy of >80 MeV. To estimate electrons energy undeflected electron beam position on Phosphor-2 was obtained from the correlation between the electron beam positions on Phosphor-1 and Phosphor-2 recorded without magnet in between. Further, posphor-3 (DRZ-high) was placed perpendicular to the laser propagation direction in the side of C-shaped magnet to record the lower energy electrons spectrum, and it covered a range of ~15 - 80 MeV. To study the laser plasma interaction and laser channeling inside plasma, shadowgraphy was performed by using the leakage of the main laser beam ($\lambda$= 800 nm) as the probe beam with variable delay in the few ps regime and imaged on to a 14 bit CCD camera with a magnification of 2X. The electron beam charge was estimated using standard calibration data available for various phosphor screens used [76-78].

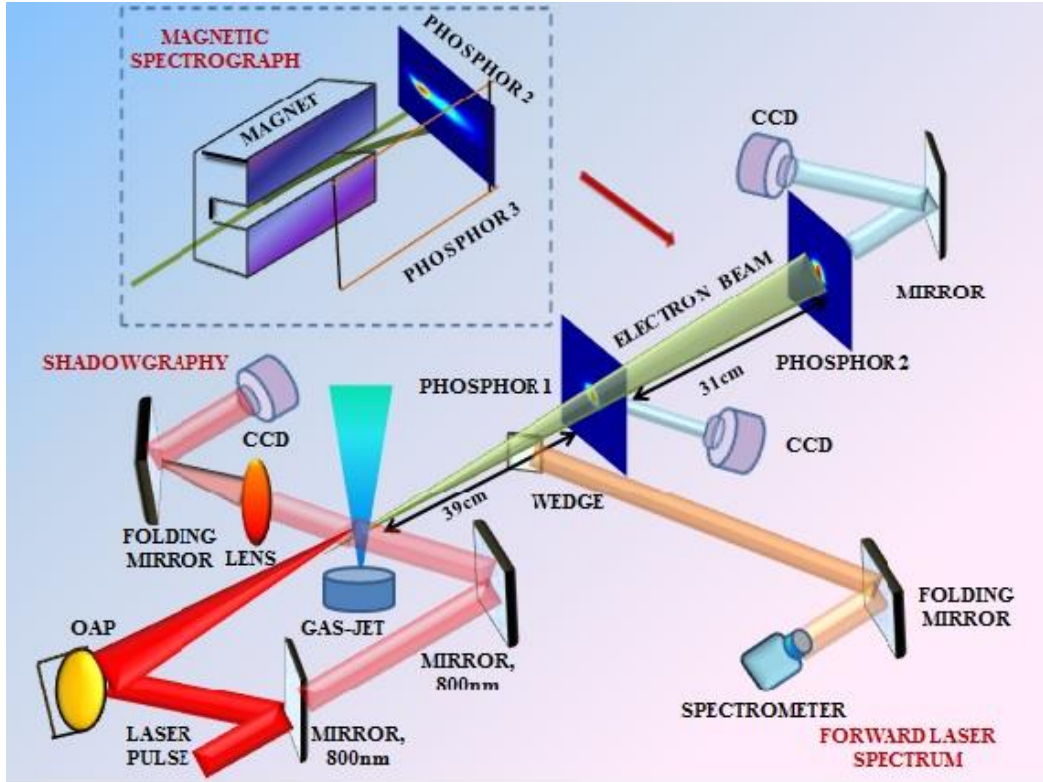

FIG.1: Schematic of the experimental set up.

## 3. Experimental Results:

The present experiment was carried out with laser pulse of ~60 - 70 fs ($a_0$ ~ 4.2 - 4.5) keeping the laser energy fixed. First high Z gas target of pure $N_2$ was used, followed by mixture of $N_2$ and He (with varying fraction). Threshold densities for electron beam generation were found to be different for different fraction of He in $N_2$ and also corresponding electron beam spectra recorded were also different. Various threshold densities observed were: ~$1.6\times10^{18}$ cm$^{-3}$ (pure $N_2$), ~$2\times10^{18}$ cm$^{-3}$ ($N_2$ + 50% He), ~$2.3\times10^{18}$ cm$^{-3}$ ($N_2$ + 58% He), ~$3.1\times10^{18}$ cm$^{-3}$ ($N_2$ + 75% He) and ~$4.3\times10^{18}$ cm$^{-3}$ ($N_2$ + 98% He) respectively. Stable generation of electron beams with high reproducibility was observed during the experiment.

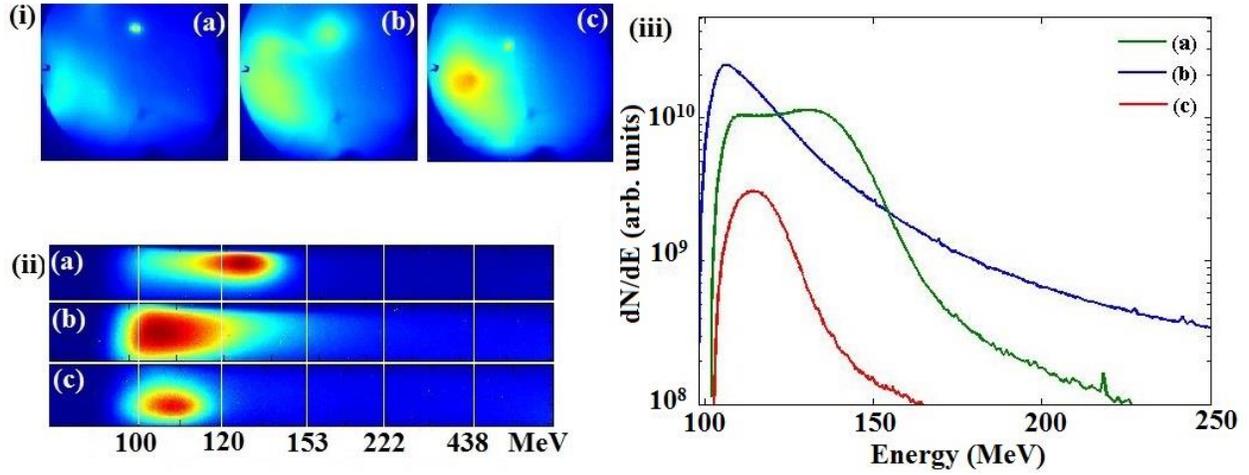

FIG.2. Experimental results: $N_2$ gas-jet target: Threshold density: $\sim 1.6 \times 10^{18}$ cm$^{-3}$: (i) (a-c) Electron beam profiles showing narrow intense electron beam with diffused background, (ii) (a-c) Corresponding dispersed raw spectra of narrow e-beam (iii) (a-c) Lineouts of corresponding spectra.

Fig. 2 (i a-c) shows a series of electron beam profiles which shows generation of directional electron beam profiles with divergence in the range of 5.8 - 6.7 mrad (Avg.: 6.2 mrad) for pure $N_2$ gas-jet target at a threshold density of $\sim 1.6 \times 10^{18}$ cm$^{-3}$. In addition to the narrow electron beam spot a large shot to shot fluctuating background was also seen. Total charge in the narrow beam varied in the range of 65 - 174 pC. Fig. 2 (ii a-c) and Fig. 2 (iii a-c) shows the corresponding dispersed electron beam and spectra respectively. The spectra showed a lower energy cut-off with an average peak of $\sim 100$ - 125 MeV, and an exponential broad distribution towards higher energy side. Maximum energy corresponding to 10% of the peak signal was in the range of 136 - 161 MeV (Avg.: 151 MeV). With further increase in the density to $\sim 3.5 \times 10^{18}$ cm$^{-3}$, similar electron beam profiles with divergence in the range of 7-10 mrad (Avg.: 8.3 mrad) were observed (Fig. 3 i a-e). Fig. 3 (ii a-e) and Fig. 3 (iii a-e) shows the corresponding dispersed electron beams and the spectra with peak energies in the range of 138 - 173 MeV (Avg.: 148 MeV) and maximum energy in the range of 167 - 248 MeV (Avg.: 209 MeV). The spectra observed are mostly quasi-thermal. The total charge in the narrow electron beam was comparatively higher and found to be ~180 – 250 pC.

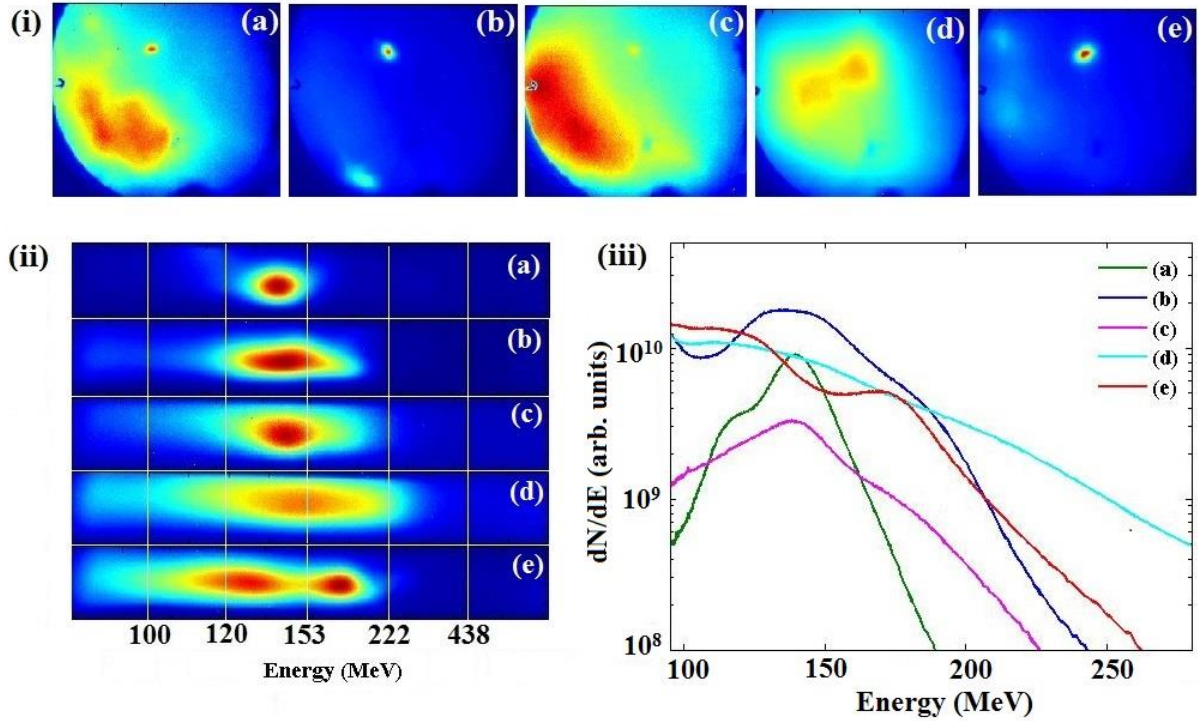

FIG.3. Experimental results: $N_2$ gas-jet target: Threshold density: ~$3.5\times10^{18}$cm$^{-3}$: (i) (a-e) Electron beam profiles showing narrow intense electron beam with diffused background, (ii) (a-e) Corresponding dispersed raw spectra of narrow e-beam (iii) (a-e) Lineouts of corresponding spectra.

Next, electron acceleration in mixed gas-target of $N_2$ + He was studied. Fig.4 (i a-j) and Fig.4 (ii a-j) show the electron beam spectra and its spectra obtained from mixed gas target of $N_2$ + 50% He for a threshold density of ~$2\times10^{18}$ cm$^{-3}$. Interestingly, generation of quasi-monoenergetic electron beams was observed with peak energies in the range of 144 - 205 MeV (Avg.: 168 MeV) and energy spread in the range of ~17% - 26% (Avg.: ~21%). Probability of generation of quasi-monoenergetic electron beams was found to be 65 - 70% in a series of consecutive shots during the experiment. Maximum energy was in the range of 171 - 265 MeV (Avg.: 206 MeV). Fig.4 (iii a-d) shows typical electron beam profiles with narrow electron beam with divergence varying in the range of 7 - 9 mrad (Avg.: 7.5 mrad) and an associated diffused background. The charge contained in the narrow electron beams was 140 - 300 pC.

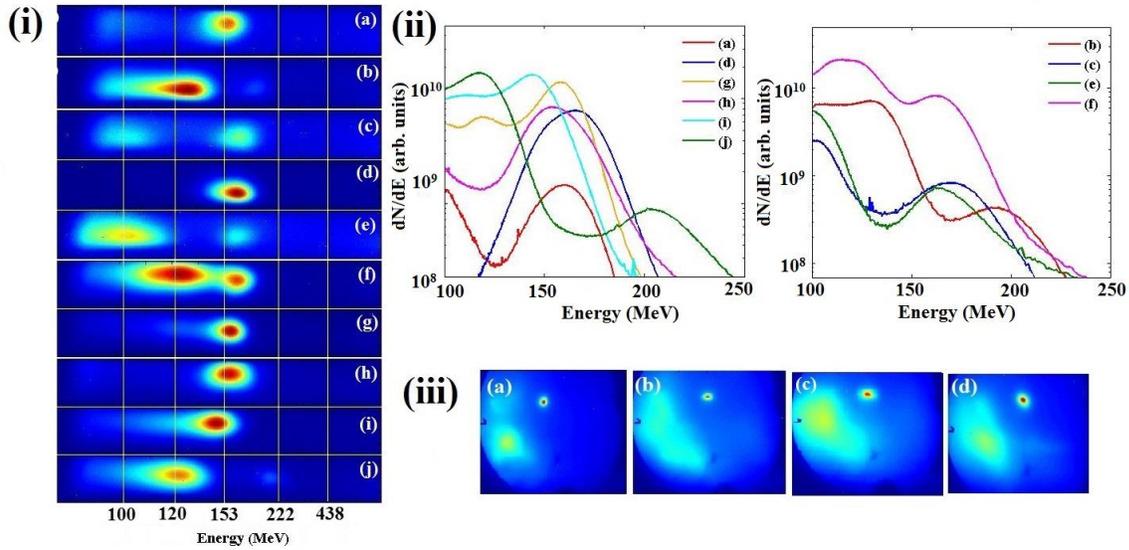

FIG.4. Experimental results: $N_2$ + 50% He gas-jet target: Threshold density: ~$2\times10^{18}$ cm$^{-3}$: (i) (a-j) Dispersed raw spectra of narrow e-beam generated (ii) (a, d, g, h, i, j) Lineouts of corresponding spectra shown in one frame and (ii) (b, c, e, f) Lineouts of corresponding spectra shown in separate frame for clarity. (iii) Electron beam profiles showing narrow intense electron beam with diffused background.

For comparison, Fig.5 (i - iii) show typical dispersed electron beams and corresponding spectra obtained from mixed gas-jet targets of $N_2$ + 58% He, $N_2$ + 75% He and $N_2$ + 98% He at corresponding threshold density of ~$2.3\times10^{18}$ cm$^{-3}$, ~$3.1\times10^{18}$ cm$^{-3}$ and ~$4.3\times10^{18}$ cm$^{-3}$ respectively. For $N_2$ + 58% He, at a threshold density of ~$2.3\times10^{18}$ cm$^{-3}$, electrons with peak energies in the range of 117 - 177 MeV (Avg.: 153 MeV) and maximum energy in the range of 147 - 213 MeV (Avg.: 184 MeV) were generated (Fig.5 i). Probability of generation of quasi-monoenergetic energy spectra was reduced to 30 - 40% of the total consecutive shots, which was almost absent for further higher percentage of He used. For $N_2$ + 75% He (Fig.5 ii), at a higher threshold density of ~$3.1\times10^{18}$ cm$^{-3}$, electron beam peak energy were observed in the range of 109 - 130 MeV (Avg.: 118 MeV) with maximum energy in the range of 140 - 155 MeV (Avg.: 147 MeV). At still higher fraction of He, i.e. $N_2$ + 98% He (Fig.5 iii), at a threshold density of ~$4.3\times10^{18}$ cm$^{-3}$, quasi-thermal electron spectrum with formation of multiple bunches of electrons were observed and maximum energy varied in the range of ~213 - 250 MeV (Avg.: 235 MeV).

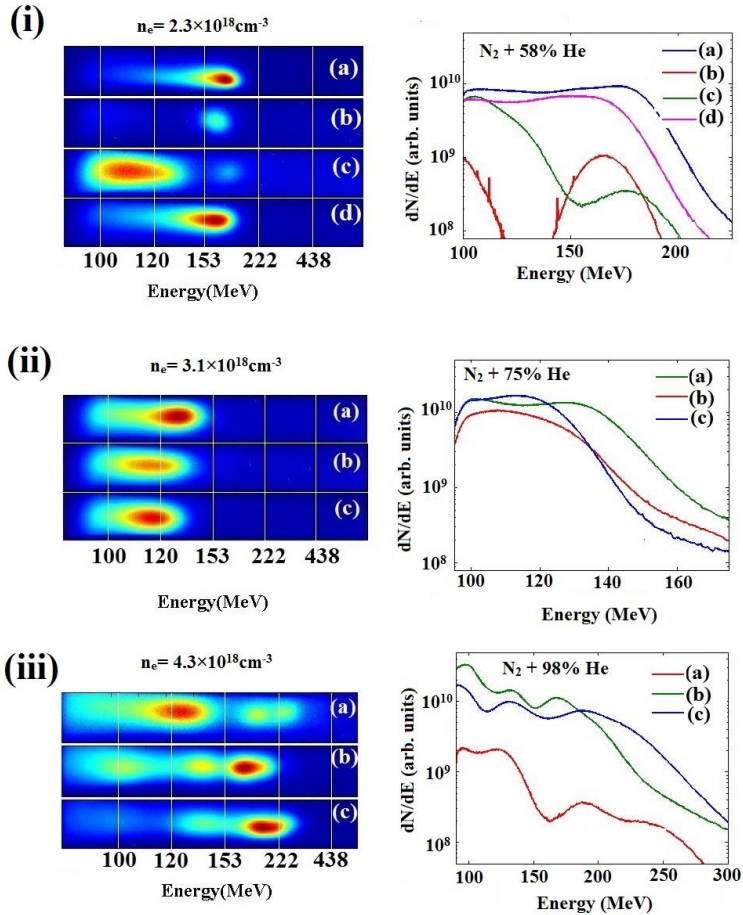

FIG.5. Experimental results: Typical dispersed raw spectra of narrow e-beam generated and corresponding spectra for (i) $N_2$ + 58% He gas-jet target: Threshold density: $\sim n_e$= $2.3\times10^{18}$cm$^{-3}$, (ii) $N_2$ + 75% He gas-jet target: Threshold density: $\sim 3.1\times10^{18}$cm$^{-3}$ and (iii) $N_2$ + 98% He gas-jet target: Threshold density: $\sim 4.3\times10^{18}$cm$^{-3}$.

In Fig.6 (i-v) shadowgrams of laser created channels inside plasma and the corresponding electron beam profiles for various targets used at corresponding threshold density are shown. Laser channeling associated with relativistic self-focusing and filamentation was observed in various targets used. This is also consistent with the corresponding electron beam profiles which showed a narrow electron beam from the main channel and a large diffused beam from the filamentation. Stable laser propagation is expected in low Z gas- targets, as also observed in case of $N_2$ + 98% He (Fig. 6-v), which is a combination also used in various other investigations on electron acceleration with ionization induced injection. However, interestingly, even for a combination of $N_2$ + 50% He target (Fig.6-ii), comparatively stable and longer channels were observed.

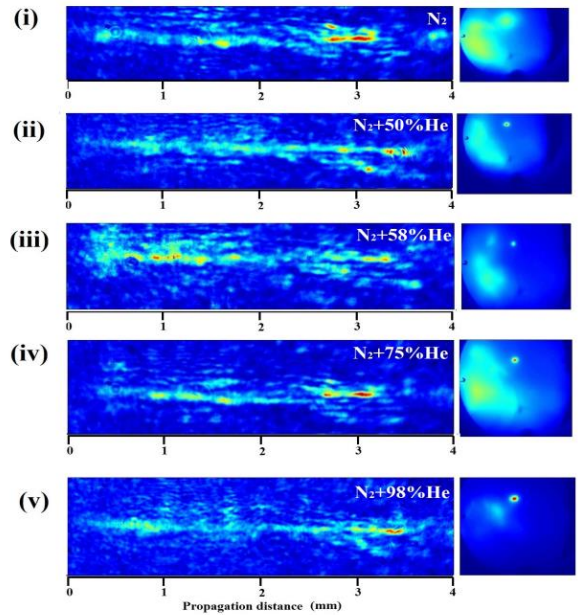

FIG.6. Experimental results: Shadowgrams of laser created channel and corresponding electron beam profiles for different gas targets (threshold densities) of (i) pure $N_2$ (~$1.6\times10^{18}$ cm$^{-3}$), (ii) $N_2$ + 50% He (~$2\times10^{18}$ cm$^{-3}$), (iii) $N_2$ + 58% He (~$2.3\times10^{18}$ cm$^{-3}$), (iv) $N_2$ + 75% He (~$3.1\times10^{18}$ cm$^{-3}$), and (v) $N_2$ + 98% He (~$4.3\times10^{18}$ cm$^{-3}$).

## 4. Discussion:

### (A) Identification of Acceleration Mechanisms:

The present experiment was carried out with laser pulse of ~60 - 70 fs ($a_0$ ~ 4.2 - 4.5) at a fixed laser energy. The laser power was varied from ~17.4 - 20.3 TW, and $P/P_c$ values, corresponding to density ~1.6 - $4.3\times10^{18}$ cm$^{-3}$, varied in the range of ~1 - 2.7, where $P_c$ (GW) =17.4($n_c/n_e$) is the critical power for self-focusing [79], $n_c$ is the critical density and $n_e$ is the electron density. For the used laser and plasma parameters, $L/\lambda_p$ varied from 0.8 - 1.12 i.e. $L\sim\lambda_p$, with laser pulse length (L) of ~18 - 21 μm, and $\lambda_p$ of ~16 -26 μm for the range of plasma density used i.e. ~1.6 - $4.3\times10^{18}$ cm$^{-3}$. In such a scenario there could be significant overlap of the laser pulse with accelerated electrons and therefore contribution of DLA mechanism of acceleration has also to be considered [30-32, 66-71].

The maximum energy of electrons estimated theoretically for linear [1] /non-linear [9] wakefield mechanism are 1026 / 1593 MeV (density~$1.6\times10^{18}$ cm$^{-3}$), 850 / 1275 MeV (density~$2\times10^{18}$ cm$^{-3}$), 739 / 1108 MeV (density~$2.3\times10^{18}$ cm$^{-3}$), 548 / 822 MeV (density~$3.1\times10^{18}$ cm$^{-3}$) and 395 / 593 MeV (density~$4.3\times10^{18}$ cm$^{-3}$). These values are much higher than the observed values at corresponding densities using variety of targets. For similar conditions energy estimated theoretically from DLA mechanism [32] are 147 MeV (density ~$1.6\times10^{18}$ cm$^{-3}$), 151 MeV (density ~$2\times10^{18}$ cm$^{-3}$), 150 MeV (density ~$2.3\times10^{18}$ cm$^{-3}$), 138 MeV (density $\sim3.1\times10^{18}$ cm$^{-3}$) and 130 MeV (density $\sim4.3\times10^{18}$ cm$^{-3}$) and are found to be close to the experimentally observed values. This suggests the possible applicability of DLA in our experimental conditions as shown in the Fig.7 where a comparison of the observed maximum electron energy with the theoretically estimated maximum electron energy from DLA and wakefield are shown. For pure $N_2$ gas-jet target, stable generation of relativistic electron beam was observed at a threshold density of $\sim1.6\times10^{18}$ cm$^{-3}$ with average peak energies of $\sim117$ MeV (Fig. 2). The spectra are mostly exponential at threshold density. However, with increase in density to $\sim3.5\times10^{18}$ cm$^{-3}$, average peak electron energy was found to increase to $\sim148$ MeV with quasi-thermal spectrum (Fig.3). The observation of exponential / quasi-thermal spectrum and increase in energy with density are the signature of applicability of DLA mechanism [30, 32]. For mixed gas-target also, electron acceleration was observed at the slightly higher density of $\sim2 - 4.3 \times10^{18}$ cm$^{-3}$, and could be attributed to pure DLA on the lower density side with increasing role of wakefield on the higher density side. It may be noted here that the above densities were much lower than the self-injection threshold density of $\sim5.8\times10^{18}$ cm$^{-3}$, observed experimentally using pure He gas-jet target for 60 fs pulse duration in the same experimental set up, in which case acceleration mechanism could be wakefield.

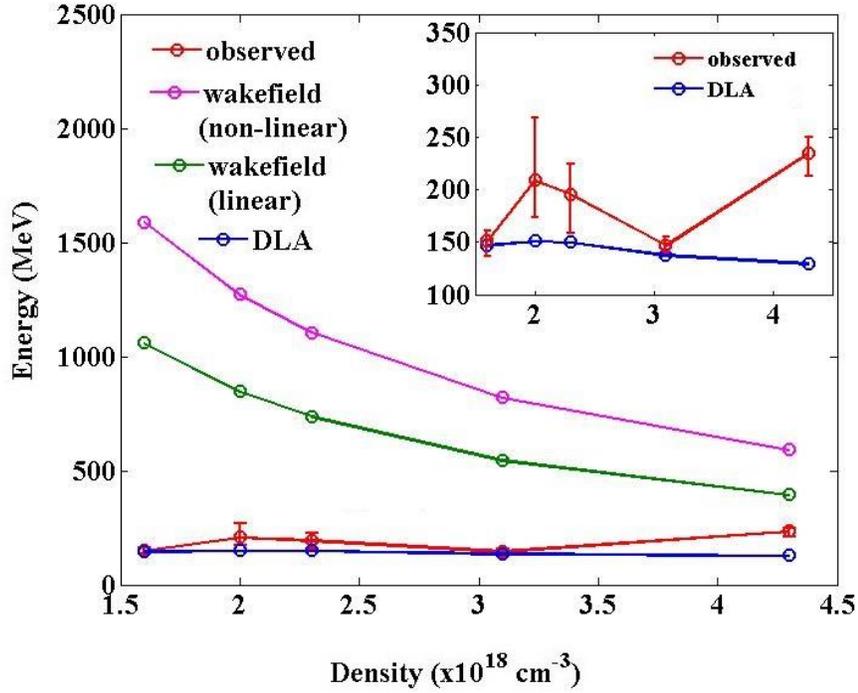

FIG.7. Comparison of experimentally observed maximum energy (red curve) of electrons with theoretical calculated maximum energy gain from wakefield (non-linear): magenta curve, wakefield (linear): green curve and DLA: blue curve, for the range of plasma density used. Inset shows the enlarged view of comparison of observed energy gain (red curve) with only DLA (blue curve). Error bars in maximum electron energy at a given density corresponds to observed shot-to-shot energy fluctuation.

**(B) 2D PIC simulation verification of acceleration mechanisms:**

Applicable acceleration mechanism was also verified by 2D PIC simulations performed using code EPOCH [74], using three different gas-jet targets of pure $N_2$ at a density of $1.2 \times 10^{18}$ cm$^{-3}$, mixed gas target of $N_2$ + 50 % He at a density of $2.6 \times 10^{18}$ cm$^{-3}$ and $N_2$ + 75% He at a density of $3.7 \times 10^{18}$ cm$^{-3}$. The normalized laser and wakefield strengths have been plotted after a propagation of ~2.6 mm inside plasma for the above conditions (Fig.8 a-d). In case of $N_2$ and mixed gas target of $N_2$ + 50% He, negligible modulation of the laser pulse is observed (Fig.8 a-b) with negligible strengths of wakefield (Fig.8 d) suggesting dominant DLA regime of acceleration. However, for mixed gas composition of $N_2$ + 75% He with higher threshold density, some modulation of the laser pulse and increase in the wakefield strengths were observed (Fig. 8 c & d). This suggests increasing role of wakefield towards higher density leading to a hybrid acceleration regime [66, 71]. The above observations are also supported by the fact that the present experimental condition of $L \sim \lambda_p$ and $P \sim P_c$ may not support strong self-modulation of the laser pulse and hence generation of strong wakefield. Next, snapshots of the electron density profiles of three different gas targets at three different propagation distances of ~0.5 mm, ~1.4 mm and ~2.6 mm are shown in Fig. 9 (i-iii) respectively. In all the three cases betatron oscillation of electrons is clearly seen in the laser created channel, suggesting acceleration by pure DLA mechanism [67, 69-71].

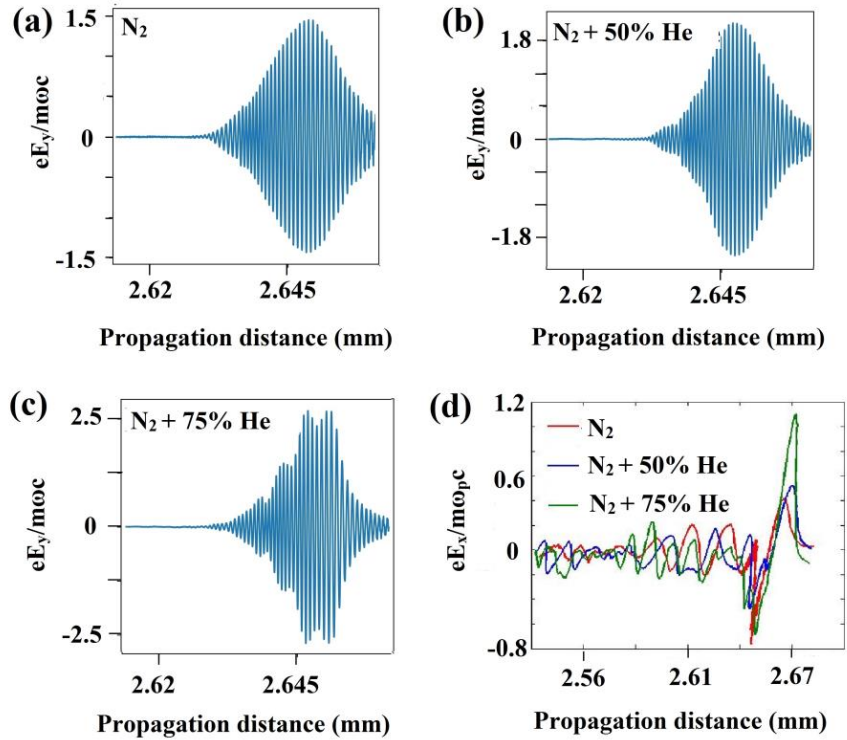

FIG.8. Simulation results: Lineouts of normalized laser field ($E_y$) after propagation of ~2.6 mm inside plasma for (a) $N_2$ gas-jet target, Threshold density: $1.2\times10^{18}$ cm$^{-3}$, (b) $N_2$ + 50% He gas-jet target, Threshold density: $2.6\times10^{18}$ cm$^{-3}$ and (c) $N_2$ + 75% He gas-jet target, Threshold density: $3.7\times10^{18}$ cm$^{-3}$ (d) Comparison of normalized wakefield ($E_x$) for the three gas-jet targets after propagation of ~2.6 mm inside plasma. Simulation parameters: Laser wavelength: 800 nm, Laser pulse duration: 60 fs, $a_0$=4.5, Laser polarization: Y-direction, Laser propagation: X-direction, Plasma length: ~2.7 mm, Simulation box size (resolution): 100 μm ($\lambda$/30) ×150 μm ($\lambda$/10) in longitudinal and transverse direction respectively.

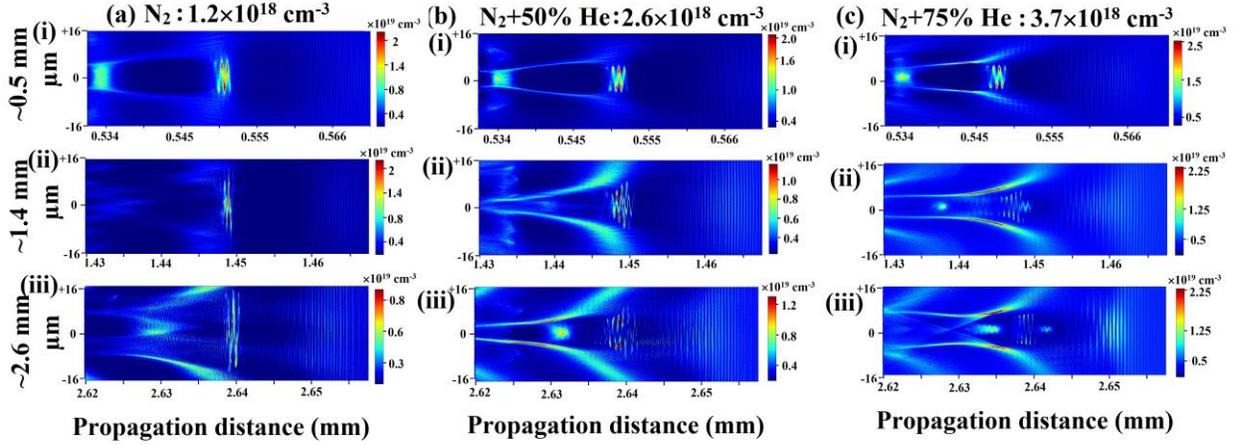

FIG.9. Simulation results: Electron density profiles for (a) $N_2$ gas-jet target, Threshold density : $1.2\times10^{18}$ cm$^{-3}$, (b) $N_2$ + 50% He gas-jet target, Threshold density: $2.6\times10^{18}$ cm$^{-3}$ and $N_2$ + 75% He gas-jet target, Threshold density: $3.7\times10^{18}$ cm$^{-3}$ for propagation length of (i) ~0.5 mm, (ii) ~1.4 mm and (iii) ~2.6 mm respectively.

**(C) Optimization of threshold plasma density through gas target composition in DLA regime:**

For the pure DLA regime in the density range of ~1.6 – 3.1×10$^{18}$ cm$^{-3}$ the maximum energy gained by electrons was observed at an optimum density of ~ 2×10$^{18}$ cm$^{-3}$ for mixed gas target of N$_2$ + 50% He (inset of Fig.7). Electron beams at threshold densities below and above this optimum value corresponding to different mixed gas targets were observed with comparatively lower energies. However, maximum electron energy was found to increase at a higher threshold density of ~4.3×10$^{18}$ cm$^{-3}$ for mixed gas target of N$_2$ + 98% He, in which case the increase in energy is attributed to the enhanced contribution of wakefield leading to the hybrid acceleration mechanism.

This observation was also supported by 2D PIC simulation (Fig.9). For an initial propagation of ~0.5 mm inside plasma (Fig.9 i a-c), in all the three cases of different mixed gas targets, betatron oscillation of a bunch of electrons in the channel is observed. On further propagation of laser inside plasma up to a distance of ~1.4 mm in case of N$_2$ (Fig.9 ii-a), electrons were found to develop enhanced oscillation amplitude with loss of synchronicity in the period of oscillation i.e. a fraction of the injected electrons started to move out of phase with the laser pulse. Whereas, in case of N$_2$ + 50% He (Fig.9 ii-b), although amplitude of oscillation of electrons increased but they still maintain synchronicity i.e. the electrons are still in phase with the laser pulse forming a bunch. However, in case of N$_2$ + 75% He (Fig.9 ii-c), a fraction of the electrons were found to oscillate in the laser field with larger oscillation amplitude and the remaining electrons were found to dephase out of the laser field and oscillate with smaller oscillation amplitude. This suggests the presence of wakefield and a hybrid regime of acceleration. On further propagation of ~2.6 mm inside plasma, electrons were found to be more dephase out of the laser field in case of $N_2$ (Fig.9 iii-a), whereas, in case of $N_2$ + 50% He, the electrons still maintain a bunch structure and remain in phase with the laser pulse suggesting a pure DLA mechanism (Fig.9 iii-b). However, in case of $N_2$ + 75% He, the electrons with lower oscillation amplitude interacting with wakefield now forms a distinct bunch structure and clearly separate from DLA bunch with lower oscillation amplitude, suggesting energy gain from both DLA and wakefield (Fig.9 iii-c). Similar kind of oscillation of electrons showing difference in the oscillation amplitude of the DLA and non-DLA accelerated electrons in a plasma channel has also been observed (c/f from Fig. 3c of ref. 69 and c/f from Fig. 4a of ref. 70), earlier in 2D PIC simulations, [69, 70].

**(D) Theoretical estimation of rate of dephasing of electrons in DLA regime:**

A theoretical analysis was also performed to support the above observed optimum value of plasma density for generation of comparatively higher energy electron beams through DLA mechanism in given experimental conditions. The electron acceleration through DLA in a plasma channel has been studied theoretically [32, 33] and effect of rate of dephasing on electron energy has been identified which could be dependent on various fields [80, 81]. Here we show variation of rate of dephasing with plasma density. Rate of dephasing (R) is defined as rate of change of phase of electrons trapped in the laser field and is given by:

$$R = -\frac{1}{\omega}\frac{d\phi}{d\tau} = -\frac{1}{\omega}\frac{d\phi}{dt}\frac{dt}{d\tau} = -\frac{\gamma}{\omega}\frac{d\phi}{dt} = -\frac{\gamma c}{\omega}\frac{d\phi}{dz} = -\gamma\frac{d\phi}{d\xi} \qquad (1)$$

where, τ is the proper time defined as $d\tau/dt = 1/\gamma$, ω is the laser frequency, ϕ is the phase of the electron with the laser field and $\xi = z\omega/c$ is the normalized distance along propagation direction of the laser.

The change of phase of electrons interacting with the laser field (ϕ) with the propagation distance (ξ) is given by [32]:

$$\frac{d\phi}{d\xi} = \frac{1-\omega_{b0}/\omega\sqrt{\gamma}}{(1-\alpha_0 - 1/\gamma^2)^{1/2}} - \eta \qquad (2)$$

Therefore,

$$R = -\gamma[\frac{1-\omega_{b0}/\omega\sqrt{\gamma}}{(1-\alpha_0 - 1/\gamma^2)^{1/2}} - \eta] \qquad (3)$$

where, $\omega_{b0} = \omega_b \sqrt{\gamma}$ represents the bounce frequency of the oscillation, $\alpha_0 = v_{xA}^2/2c^2$ where $v_{xA}$ is the on axis velocity of electrons, $\eta = (1 - \omega_p^2/\omega^2 (1+a_0^2/2)^{-1/2})^{1/2}$ is the ratio of group velocity of the laser in plasma to that in vacuum which also gives a measure of electron density for a given $a_0$. Eq. (3) shows that η i.e. density is a crucial factor which decides the rate of dephasing of electrons. The first term inside the bracket of right hand side of Eq. (3) denotes the energy gain of electrons oscillating with betatron frequency $\omega_b$ in the laser field and the second term denotes density, and hence, R i.e. rate of dephasing is the interplay between density and energy gain. In Fig.10-a, we plot the variation of R with density for γ=γ$_{opt}$ (optimum energy for electrons to be trapped and in phase with the laser field at corresponding density) for the case of 60 fs. It shows that R is minimum towards lower density values of 1.6 - 2×10$^{18}$ cm$^{-3}$ which suggests that electrons interact maximum with the laser field and gain maximum energy. This is consistent both with the experimental observations and simulation results showing optimum density of ~$2\times10^{18}$ cm$^{-3}$.

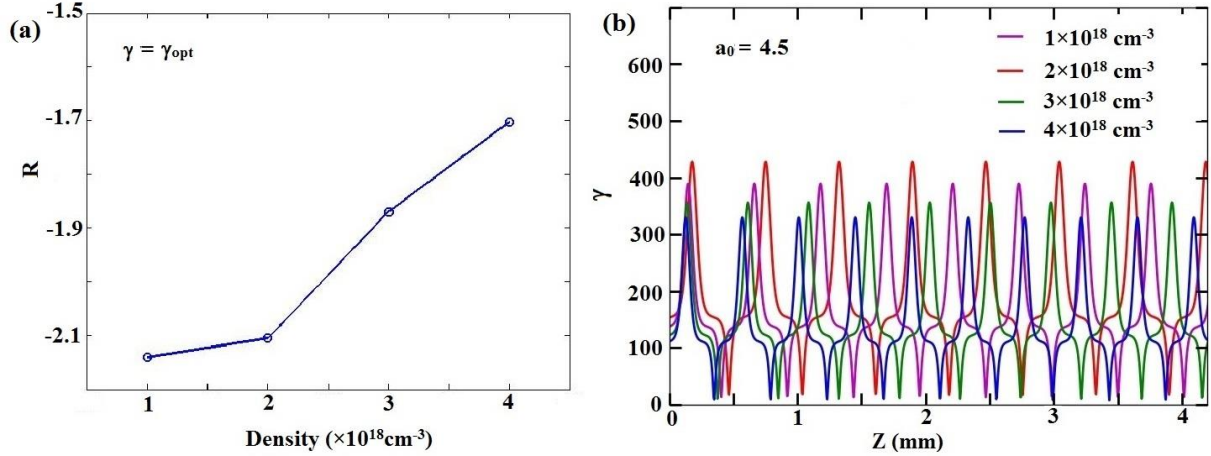

FIG.10. Theoretical results: (a) Plot of dephasing rate (R) vs density and (b) Plot of energy gain of electrons (γ) with propagation distance (z) for different densities of the present experimental conditions.

In Fig.10-b, we plot the variation of energy gain γ with propagation distance z for the range of densities used in the present experiment. It is observed that for the present density range all electrons are trapped in the laser field, however, particularly at a density of $2\times10^{18}$cm$^{-3}$ (red curve) the dephasing rate of electron is slower compared to the other densities. Within the 4 mm length of laser propagation, electrons at density of $2\times10^{18}$ cm$^{-3}$ covers ~7 complete phase space rotation whereas at other densities it covers ~8 complete phase space rotations. This suggests that electrons at density of $2\times10^{18}$cm$^{-3}$ remains in phase with the laser field for a longer period of time compared to that at other densities leading to maximum energy gain.

**(E) Plausible explanation of generation of quasi-monoenergetic electron beams:**

Next, at the optimum composition of $N_2$ + 50% He (~$2\times10^{18}$ cm$^{-3}$) for higher energy electron beams, generation of quasi-monoenergetic electron beams were also observed (Fig. 4) in almost 70% of the shots. Generation of quasi-monoenergetic electron beams in a hybrid (wakefield + DLA) regime of acceleration has been reported earlier in a PIC simulation study and it was suggested that DLA accelerated electrons could also have lower energy spread [69, 70]. The theoretical analysis above also showed that for an optimum density of $2\times10^{18}$ cm$^{-3}$, electrons interact longer with the laser field which could be favorable for higher energy gain, electron bunching and appearance of quasi-monoenergetic feature. Such bunching of electrons were also seen in the simulations, and found to be more pronounced for a density close to the case of $N_2$ + 50% He (Fig.9 b i-iii).

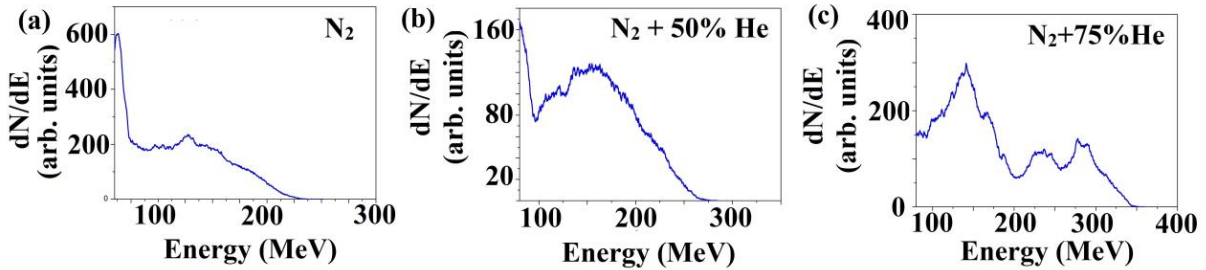

FIG.11. Simulation results: Electron energy spectrum (a) $N_2$ gas-jet target, Threshold density: $1.2\times10^{18}$ cm$^{-3}$, (b) $N_2$ + 50% He gas-jet target, Threshold density: $2.6\times10^{18}$ cm$^{-3}$ and (c) $N_2$ + 75% He gas-jet target, Threshold density: $3.7\times10^{18}$ cm$^{-3}$ after propagation of ~2.6 mm inside plasma.

Further, electron spectra generated through 2D PIC simulations also showed quasi-monoenergetic feature. Electron energy spectrum generated through simulations for the present experimental conditions after a propagation distance of ~2.6 mm are shown in Fig. 11 (a-c) for the three different gas targets respectively. At the optimum mixed gas target of $N_2$ + 50% He at a threshold density of ~$2.6\times10^{18}$ $cm^{-3,}$ the spectra shows quasi-monoenergetic feature with peak energy of ~160 MeV and maximum energy of ~220 MeV (Fig. 11-b). The increase in energy and the quasi-monoenergetic feature observed in this case suggests the effect of lower dephasing rate and hence longer interaction with the laser pulse as shown in Fig.9 b (i-iii). Whereas, in case of $N_2$, at a lower threshold density of $1.2\times10^{18}$ $cm^{-3}$, electron energy spectra is found to be quasi-thermal with maximum energy of ~180 MeV (Fig. 11 a). This is due to early dephasing of electrons with the laser field as observed in Fig.9 a (i-iii). On the other hand, in case of mixed gas target of $N_2$ + 50% He, at a higher threshold density of $3.7\times10^{18}$ $cm^{-3}$, the spectra is also quasi-thermal with multiple peak formation with maximum electron energy of ~310 MeV (Fig. 11 c). Formation of multiple peaks in the energy spectrum is the manifestation of the two different electron bunch formations with different oscillation amplitude accelerated by DLA and wakefield (Fig.9 c i-iii).

## 5. Conclusion:

In conclusion, an experimental investigation on electron acceleration in $N_2$ and $N_2$ + He mixed gas targets using Ti:sapphire laser pulses of 60-70 fs duration was performed to study role of DLA. It was found that varying doping concentrations of He in $N_2$ gas target lead to change in the threshold density for generation of relativistic electron beams. Threshold density was found to increase from ~1.6 (pure $N_2$) to $4.3\times10^{18}$ $cm^{-3}$ for 98% He. The varying threshold density in turn governed the electron beam properties. An optimum was achieved at a composition of $N_2$ + 50% He (threshold density of ~$2\times10^{18}$ cm$^{-3}$), where generation of quasi-monoenergetic electron beams with average peak energy of ~168 MeV and average energy spread of ~21% were observed. Electron acceleration was attributed to pure DLA mechanism towards the lower density side and with increasing role of wakefield on the higher density side leading to hybrid regime of acceleration. A theoretical analysis was performed to estimate the rate of dephasing of electrons with the laser field with varying plasma density and could be used to explain the observation of an optimum composition of gas target i.e. fraction of He in $N_2$ (in turn threshold plasma density). Further, 2D PIC simulations were performed using code EPOCH to support the experimental observations. Such a study of optimization of electron beam properties by varying He gas mixture compositions in a high-Z $N_2$ gas target in a DLA dominated regime has not been reported earlier. Such a regime of electron acceleration with generation of high-energy, high-charge electron beams through DLA would be beneficial for development of high flux betatron radiation source for various practical applications.

**Acknowledgements:**

The authors would like to acknowledge the support provided by R. A. Khan, A. Singla and Sunil Meena for the laser operation, D. Karmakar for help in setting up the experiment, and R. P. Kushwaha, S. Sebastin, L. Kisku and K. C. Parmar for providing mechanical / workshop support.